\documentclass[prd,amsmath,amssymb,preprintnumbers,superscriptaddress,twocolumn,10pt,nofootinbib,showkeys,showpacs]{revtex4-1}

\pdfoutput=1
\usepackage{graphicx}
\usepackage{subfigure}

\usepackage{dcolumn}
\usepackage{bm}
\usepackage{amssymb}
\usepackage{latexsym}
\usepackage{booktabs}
\usepackage{amsmath}
\usepackage{multirow}
\usepackage{url}
\usepackage{changes}
\usepackage{soul}
\usepackage{color, xcolor}
\usepackage[colorlinks=true, linkcolor=red, citecolor=blue]{hyperref}
\usepackage{tabularx}
\usepackage{makecell}
\IfFileExists{orcidlink.sty}{\usepackage{orcidlink}}{}

\providecommand{\orcidlink}[1]{%
  \href{https://orcid.org/#1}{%
    \includegraphics[height=1.6ex]{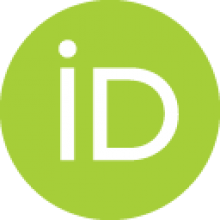}%
  }%
}

\usepackage[normalem]{ulem}
\usepackage{array}
\usepackage{enumerate}

\def\be{\begin{equation}}
\def\ee{\end{equation}}

\def\bea{\begin{eqnarray}}
\def\eea{\end{eqnarray}}
\newcommand{\bes}{\begin{equation*}}
\newcommand{\ees}{\end{equation*}}
\newcommand{\beqa}{\begin{eqnarray}}
\newcommand{\eeqa}{\end{eqnarray}}

\begin{document}

\title{Model-independent $H_0$ from GWTC-4 standard sirens and TDCOSMO 2025 strong lensing time delays}

\author{En-Yu Xiong\orcidlink{0009-0002-5269-1942}}\thanks{These authors contributed equally to this paper.}
\affiliation{Liaoning Key Laboratory of Cosmology and Astrophysics, College of Sciences, Northeastern University, Shenyang 110819, China}

\author{Ji-Yu Song\orcidlink{0009-0003-8111-0470}}\thanks{These authors contributed equally to this paper.}
\affiliation{Liaoning Key Laboratory of Cosmology and Astrophysics, College of Sciences, Northeastern University, Shenyang 110819, China}

\author{Jing-Fei Zhang\orcidlink{0000-0002-3512-2804}}
\affiliation{Liaoning Key Laboratory of Cosmology and Astrophysics, College of Sciences, Northeastern University, Shenyang 110819, China}

\author{Xin Zhang\orcidlink{0000-0002-6029-1933}}\thanks{Corresponding author: \href{mailto:zhangxin@neu.edu.cn}{zhangxin@neu.edu.cn}}
\affiliation{Liaoning Key Laboratory of Cosmology and Astrophysics, College of Sciences, Northeastern University, Shenyang 110819, China}
\affiliation{MOE Key Laboratory of Data Analytics and Optimization for Smart Industry, Northeastern University, Shenyang 110819, China}
\affiliation{National Frontiers Science Center for Industrial Intelligence and Systems Optimization, Northeastern University, Shenyang 110819, China}

\begin{abstract}

The significant discrepancy between early- and late-Universe measurements of the Hubble constant, known as the Hubble tension, remains one of the most pressing open questions in cosmology. Since both sides of the tension rely on model-dependent assumptions or multi-rung calibration chains, a cosmological-model-independent measurement of $H_0$ is essential to arbitrate this discrepancy.
In this work, we combine 142 gravitational-wave standard siren events from the Fourth Gravitational-Wave Transient Catalog with the latest TDCOSMO2025 time-delay strong lensing data to constrain $H_0$ in a cosmological-model-independent framework based on the distance sum rule.
Under the FullPop-4.0 population model with the TDCOSMO2025-only lensing configuration, we obtain $H_0 = 83.78^{+12.53}_{-10.23}\ {\rm km\,s^{-1}\,Mpc^{-1}}$, with a relative precision of $13.58\%$.
We find that the $H_0$ precision is governed primarily by the mass-sheet transformation treatment on the strong-lensing side: replacing the conservative TDCOSMO2025 hierarchical framework with the H0LiCOW method tightens the constraint to $H_0 = 75.42^{+3.74}_{-4.66}\ {\rm km\,s^{-1}\,Mpc^{-1}}$, with a relative precision of $5.57\%$.
At the current precision, all results are consistent with both the Planck and SH0ES values, and future improvements from more high-redshift dark siren events and more time-delay lens systems are expected to strengthen this model-independent approach.

\end{abstract}

\keywords{Cosmology, Hubble constant, Gravitational waves, Standard sirens, Strong gravitational lensing}


\maketitle

\section{Introduction}\label{sec:Introduction}

Over the past two decades, the $\Lambda$CDM model has served as the standard cosmological model owing to its remarkable success in explaining a wide range of cosmological observations.
However, a significant discrepancy known as the Hubble tension has emerged between early- and late-Universe measurements of the Hubble constant: the global fit of the Planck cosmic microwave background (CMB) data in the $\Lambda$CDM model yields $H_0 = 67.66 \pm 0.42\ {\rm km\,s^{-1}\,Mpc^{-1}}$~\cite{Planck:2018vyg}, while the latest distance-ladder measurement calibrated with Cepheid variables observed by JWST and HST gives $H_0 = 73.18 \pm 0.88\ {\rm km\,s^{-1}\,Mpc^{-1}}$~\cite{Riess:2025chq}, corresponding to a discrepancy at the $\sim 6\sigma$ level.
Extensive investigations of possible observational systematics have been unable to account for this discrepancy, suggesting that it may point to new physics beyond the $\Lambda$CDM model~\cite{Verde:2019ivm,Knox:2019rjx,Riess:2021jrx,Abdalla:2022yfr,DiValentino:2021izs,Perivolaropoulos:2021jda,Schoneberg:2021qvd,Kamionkowski:2022pkx,Guo:2018ans,Li:2024qso,Li:2024qus,Pang:2025lvh,Li:2025owk,Du:2025iow,Feng:2025mlo,Li:2025eqh,Li:2025dwz,Du:2025xes,Li:2025htp,Wu:2025vfs,Zhang:2025dwu,Li:2025muv,Cai:2025mas,Li:2025vuh,Wang:2021kxc,Zhao:2022bpd,Ma:2007av,Guo:2018gyo,Feng:2019jqa,Gong:2021jgg,Zhang:2014ifa,Zhu:2023gmx,Wang:2016tsz,Guo:2015gpa,Zhang:2014nta,Zhang:2007bi,Cui:2009ns,Feng:2017nss,Fu:2011ab,Feng:2017usu}.
Meanwhile, the Dark Energy Spectroscopic Instrument (DESI) baryon acoustic oscillation (BAO) data, combined with Planck CMB observations, reveal a $2.6\sigma$--$3.1\sigma$ preference for dynamical dark energy with a phantom-crossing behavior across Data Release 1~\cite{DESI:2024mwx} and Data Release 2~\cite{DESI:2025zgx}, reaching $2.8$--$4.2\sigma$ when Type Ia supernovae data are included~\cite{DESI:2025zgx}.
However, under such dynamical dark energy models, the $H_0$ value inferred from the CMB becomes even lower than the $\Lambda$CDM result, further exacerbating rather than alleviating the Hubble tension~\cite{DESI:2024mwx,DESI:2025zgx}.

The Hubble tension is fundamentally difficult to adjudicate because neither side of the discrepancy relies on a direct measurement of the local expansion rate. On the early-Universe side, the $H_0$ value inferred from the CMB is not a direct measurement but rather a model-dependent derived quantity: it shifts when the underlying cosmological model changes, and as the DESI-preferred dynamical dark energy scenario illustrates, it shifts in the wrong direction, moving further away from, rather than toward, the distance-ladder value. Similarly, BAO observations measure distances only as ratios to the sound horizon at the drag epoch, $r_d$, and in the inverse distance ladder approach, $r_d$ is calibrated from CMB data under the assumption of standard pre-recombination physics, so the resulting $H_0$ inherits the same model dependence~\cite{Verde:2019ivm,Abdalla:2022yfr,Du:2025csv}. On the late-Universe side, distance-ladder measurements rely on a multi-rung astrophysical calibration chain, each step of which carries its own systematic uncertainties. In this context, what is needed is a measurement of $H_0$ that is independent of both early-Universe cosmological model assumptions and the distance-ladder calibration hierarchy, and can serve as a genuine third-party arbiter of the tension. If such a measurement favors the CMB-inferred value, it would call into question the reliability of distance-ladder measurements; conversely, if it agrees with the distance-ladder value, it would suggest that current cosmological models have yet to identify the correct resolution to the tension.

To this end, Collett et al.~\cite{Collett:2019hrr} proposed a cosmological-model-independent method within the Friedmann-Lema\^itre-Robertson-Walker (FLRW) framework, combining strong gravitational lensing time delays with Type Ia supernovae to jointly constrain $H_0$ and the spatial curvature parameter $\Omega_K$. Specifically, the time-delay distance in a strong lensing system can be expressed as a combination of three angular diameter distances, and can be further rewritten in terms of dimensionless comoving distances, with an explicit dependence on $H_0$~\cite{Refsdal:1964blz,Suyu:2009by,Collett:2019hrr}. In the FLRW geometry, these three dimensionless distances satisfy the distance sum rule involving $\Omega_K$~\cite{Rasanen:2014mca,Collett:2019hrr,Wu:2024faw}. Therefore, by calibrating them with the distance-redshift relation of Type Ia supernovae, one can constrain $H_0$ and $\Omega_K$. However, unanchored Type Ia supernovae only provide relative distances; obtaining absolute distances requires calibration of their absolute magnitude $M_B$, which is strongly degenerate with $H_0$, thereby introducing significant uncertainty into any inference of $H_0$ via this approach~\cite{Cao:2021zpf,Song:2025ddm}.

Subsequently, this method was extended to quasars as alternative distance indicators, using either the ultraviolet-X-ray luminosity correlation~\cite{Wei:2020suh} or ultracompact radio structure measurements~\cite{Qi:2020rmm} to calibrate distances in strong lensing systems.
Qi et al.~\cite{Qi:2022sxm} further forecasted the constraining power of strongly lensed Type Ia supernovae on $H_0$ and $\Omega_K$.
These studies demonstrate that combining strong lensing time delays with independent distance indicators provides a viable route toward late-Universe cosmological-model-independent measurements, although the achievable precision still depends on the absolute calibration capability of the adopted distance indicator.

Gravitational-wave (GW) standard sirens provide an absolute distance indicator that does not rely on the traditional distance ladder, as the waveform amplitude of compact binary coalescences directly encodes the luminosity distance~\cite{Schutz:1986gp,Holz:2005df,Chen:2017rfc,Zhao:2020ole,Wang:2018lun,Zhang:2018byx,Zhang:2019ylr,Zhang:2019loq,Wang:2019tto,Jin:2020hmc,Qi:2021iic,Jin:2021pcv,Wang:2021srv,Wu:2022dgy,Jin:2022tdf,Jin:2022qnj,Wang:2022oou,Jin:2023tou,Yu:2023ico,Jin:2023sfc,Han:2023exn,Jin:2023zhi,Feng:2024mfx,Feng:2024lzh,Han:2024sxm,Dong:2024bvw,Xiao:2024nmi,Han:2025fii,Feng:2025wbz,Xiao:2025mcg,Dong:2025ikq,Zhang:2025yhi,Jin:2025dvf,Zhao:2019gyk,Song:2025bio,Song:2026kii,Zhu:2026gsu,Zhao:2026uvn,Das:2026glo,Liu:2026dug}.
For bright siren events with identified electromagnetic counterparts, the source redshift can be directly measured from the host galaxy; however, the only such event widely used in cosmology so far is GW170817~\cite{LIGOScientific:2017adf,LIGOScientific:2017apx,LIGOScientific:2017pwl}.
For the majority of GW events without electromagnetic counterparts, redshift information can be inferred statistically through galaxy catalog matching or source-population modeling, known as the dark siren method.
Although the redshift uncertainty of an individual dark siren is relatively large, dark sirens have larger event numbers and broader redshift coverage, making them a statistically powerful cosmological probe~\cite{Schutz:1986gp,DelPozzo:2011vcw,Chen:2017rfc,Gray:2019ksv,Mastrogiovanni:2023emh,Song:2025ddm,Zhao:2022yiv,Zhang:2023pqs,Zhang:2023gye,Sun:2024huw,Zhang:2024rra,Sun:2025cdq,Li:2023gtu,Du:2025odq,Song:2022siz}.

Motivated by this idea, Cao et al.~\cite{Cao:2021zpf} proposed replacing Type Ia supernovae with GW standard sirens to calibrate the distance quantities in strong lensing systems, thereby avoiding the degeneracy between the Type Ia supernova absolute magnitude and $H_0$.
Their method reconstructs the continuous distance-redshift relation using the absolute luminosity distances provided by GW standard sirens, and combines this reconstruction with strong lensing time-delay distances and the FLRW distance sum rule to jointly constrain $H_0$ and $\Omega_K$ without assuming a specific cosmological model.
However, because the number of currently observed bright sirens is still very limited, early studies mainly relied on simulated data from future GW detectors to assess the potential of this method.

Song et al.~\cite{Song:2025ddm} subsequently extended this framework to real GWTC-3 dark siren data, combining them with the strong lensing system RX J1131-1231 to obtain a cosmological-model-independent constraint on $H_0$.
This work showed that dark sirens can effectively extend the limited redshift range of bright sirens and provide observationally driven distance calibration for strong lensing systems.
Nevertheless, the GWTC-3 sample is still limited in size, and the reconstruction of the distance-redshift relation remains affected by statistical uncertainties and population-model assumptions~\cite{Mastrogiovanni:2023emh,Song:2025ddm}.
With the release of the Fourth Gravitational-Wave Transient Catalog (GWTC-4), the number of available GW standard sirens has increased significantly, providing a new opportunity to improve the precision of cosmological-model-independent $H_0$ constraints~\cite{LIGOScientific:2025jau}.

On the strong-lensing side, one of the main sources of systematic uncertainty in time-delay cosmography is the mass-sheet transformation (MST), or equivalently the mass-sheet degeneracy (MSD)~\cite{Schneider:2013sxa,Birrer:2020tax}.
The MST can rescale the time-delay distance while leaving the observed lensing image configuration nearly unchanged, thereby directly affecting the inference of $H_0$.
In recent years, the TDCOSMO collaboration has developed a hierarchical Bayesian framework that incorporates stellar kinematics of lens galaxies and external non-time-delay lens samples to constrain the population of lens mass profiles, thereby mitigating the impact of the MST on $H_0$ inference.
The latest TDCOSMO2025 analysis further includes high signal-to-noise ratio stellar kinematic data obtained with JWST, Keck, and VLT, and considers SLACS and SL2S external lens samples in its full strong-lensing analysis to improve the constraining power on $H_0$ from the lensing side~\cite{Birrer:2020tax,TDCOSMO:2025dmr}.

In this work, we present a cosmological-model-independent constraint on $H_0$ from the combination of GWTC-4 GW standard sirens and TDCOSMO2025 strong gravitational lensing time-delay (SGLTD) data.
Compared with the previous GWTC-3 analysis, the GWTC-4 sample increases the number of usable standard siren events from 47 to 142, significantly strengthening the calibration of the distance-redshift relation.
We reconstruct this relation using a third-order polynomial and connect the GW-calibrated distances with the SGLTD distance through the distance sum rule in the FLRW metric.
We also compare the FullPop-4.0 and MLTP population models, and investigate the impact of different lensing-side treatments on the results.
In particular, we examine whether the absence of external non-time-delay lens samples limits the final $H_0$ precision through insufficient constraints on MST-related parameters.

This paper is organized as follows.
Section~\ref{sec:method} introduces the cosmological-model-independent distance reconstruction method, the GW dark siren likelihood, and the SGLTD framework.
Section~\ref{sec:data} describes the GWTC-4 standard siren sample and the TDCOSMO2025 strong lensing data used in this work.
Section~\ref{sec:results} presents the joint constraints and discusses the impact of the population model, MST treatment, and external lens samples on $H_0$ inference.
Finally, Section~\ref{sec:conclusions} summarizes our main conclusions.

\section{Methodology}\label{sec:method}

\subsection{Strong-lensing time delay}

In an SGLTD system, light from a background source is deflected by the gravitational potential of the foreground lens galaxy and forms multiple images.
Because these light rays pass through different gravitational potentials and path lengths, there is a time delay between the arrival times of different images.
We denote the arrival times of two images, labeled $i$ and $j$, as $t_i$ and $t_j$, respectively.
The time delay between these two images is defined as $\Delta t_{ij} \equiv t_j - t_i$, and the corresponding Fermat potential difference is denoted as $\Delta\phi_{ij}$.
These quantities satisfy the following relation~\cite{Refsdal:1964blz,Suyu:2009by}:
\begin{equation}
\Delta t_{ij} = \frac{D_{\Delta t}}{c} \Delta\phi_{ij} = \frac{1}{H_0} \frac{d_l d_s}{d_{ls}} \Delta\phi_{ij},
\label{eq:time_delay_distance}
\end{equation}
where $D_{\Delta t}$ is the time-delay distance, and $d_l\equiv d(0,z_l)$, $d_s\equiv d(0,z_s)$, and $d_{ls}\equiv d(z_l,z_s)$ are the dimensionless distances from the observer to the lens, from the observer to the source, and from the lens to the source, respectively.

In practice, lensing observables such as image positions, image shapes, magnification ratios, and flux ratios cannot distinguish between the original projected mass distribution $\kappa(\boldsymbol{\theta})$ and the mass distribution transformed by a mass-sheet transformation (MST). The MST rescales the projected mass distribution of the lens galaxy as
\begin{equation}
\kappa_\lambda(\boldsymbol{\theta}) = \lambda \kappa(\boldsymbol{\theta}) + (1 - \lambda),
\end{equation}
which gives rise to the so-called mass-sheet degeneracy (MSD)~\cite{Suyu:2012aa,Suyu:2013kha}. MST-related uncertainties arise, on the one hand, from uncertainties in the mass profile of the deflector galaxy itself, corresponding to the internal MST. Combining lensing imaging observations with stellar kinematic measurements of the deflector helps to break or mitigate this internal MST degeneracy~\cite{Schneider:2013sxa,H0LiCOW:2019xdh}. On the other hand, large-scale structures along the line of sight can also induce an effective external convergence, which affects the propagation of light and modifies the time-delay distance inferred from lens modeling. This effect can be corrected by introducing the external convergence $\kappa_{\rm ext}$, yielding the true time-delay distance:
\begin{equation}
D_{\Delta t}^{\rm true}
= \frac{D_{\Delta t}^{\rm model}}{1-\kappa_{\rm ext}},
\end{equation}
where $D_{\Delta t}^{\rm model}$ denotes the time-delay distance inferred without correcting for the line-of-sight mass distribution~\cite{H0LiCOW:2019pvv}. In this work, we use the \texttt{hierArc} framework\footnote{\url{https://github.com/sibirrer/hierarc}} to implement the hierarchical Bayesian inference framework for the strong-lensing likelihood.

\subsection{The distance sum rule}

Under the assumption of a homogeneous and isotropic universe, the spacetime is described by the FLRW metric
\begin{equation}
ds^{2} = -c^{2} dt^{2} + a^{2}(t) \left[ \frac{dr^{2}}{1 - K r^{2}} + r^{2}d\Omega^{2} \right],
\end{equation}
where $c$ is the speed of light, $K$ is the spatial curvature constant, and $a(t)$ is the cosmic scale factor that characterizes the expansion of the Universe as a function of cosmic time $t$. The Hubble parameter is defined as $H \equiv \dot{a}/a$, with its present value denoted by $H_{0}$. For a strong gravitational lensing system, the angular diameter distance $D_{A}(z_{l}, z_{s})$ between the lens at redshift $z_{l}$ and the source at redshift $z_{s}$ can be expressed in terms of the dimensionless comoving distance $d(z_{l}, z_{s})$ as
\begin{equation}
d(z_{l}, z_{s}) = (1 + z_{s})\, H_{0}\, D_{A}(z_{l}, z_{s})/c,
\end{equation}
which can be written as
\begin{equation}
d(z_{l}, z_{s}) = \frac{1}{\sqrt{|\Omega_{K}|}}\,
\mathrm{sinn}\!\left[ \sqrt{|\Omega_{K}|}
\int_{z_{l}}^{z_{s}} \frac{H_{0}}{H(z)}\, dz \right],
\end{equation}
where
\begin{equation}
\mathrm{sinn}(x) =
\begin{cases}
\sin(x), & \Omega_{K} < 0, \\
x, & \Omega_{K} = 0, \\
\sinh(x), & \Omega_{K} > 0.
\end{cases}
\end{equation}
Here $\Omega_K \equiv -Kc^2/H_0^2$ is the present-day curvature density parameter, with $a_0=1$.

Therefore, under the FLRW metric, the three dimensionless distances $d_l$, $d_s$ and $d_{ls}$ in a strong lensing system satisfy the distance sum rule~\cite{Rasanen:2014mca}:
\begin{equation}
\frac{d_{ls}}{d_s} = \sqrt{1+\Omega_K d_l^2} - \frac{d_l}{d_s}\sqrt{1+\Omega_K d_s^2},
\end{equation}
which can also be rewritten as
\begin{equation}
\frac{d_l d_s}{d_{ls}} =
\frac{1}{
\sqrt{1/d_l^2 + \Omega_K}
-
\sqrt{1/d_s^2 + \Omega_K}
}.
\label{eq:distance_sum_rule}
\end{equation}

The combination $d_l d_s/d_{ls}$ appearing in Eq.~\eqref{eq:distance_sum_rule} is directly related to the time-delay distance $D_{\Delta t}$ in SGLTD systems.
According to Eq.~\eqref{eq:time_delay_distance}, the time delay can be written as $\Delta t_{ij} = (1/H_0)(d_l d_s/d_{ls})\Delta\phi_{ij}$, where the time delay $\Delta t_{ij}$ and the Fermat potential difference $\Delta\phi_{ij}$ are determined by the lensing observations.
If $d_l$ and $d_s$ can be independently calibrated using GW standard sirens, the distance sum rule determines $d_l d_s/d_{ls}$ for a given $\Omega_K$, leaving $H_0$ as the only remaining unknown.
This procedure constrains $H_0$ without assuming a specific form of $H(z)$, relying only on the validity of the FLRW metric.

\subsection{GW dark standard siren method}

As introduced in Section~\ref{sec:Introduction}, GW standard sirens directly measure the luminosity distance $D_L$ from the waveform amplitude without relying on the cosmic distance ladder. For events without identified electromagnetic counterparts, the dark siren method infers redshifts statistically by cross-matching the GW three-dimensional localization volume with galaxy catalogs~\cite{Schutz:1986gp,Mastrogiovanni:2023emh,Gray:2023wgj}. Specifically, one constructs a statistical redshift distribution for each event using information such as candidate host galaxy positions, luminosities, and catalog completeness.
In addition, the source-frame population model provides complementary redshift information through the spectral siren effect: since the observed detector-frame masses are redshifted relative to the source-frame masses, a well-constrained mass distribution, particularly one exhibiting distinct features such as peaks or gaps, helps break the mass-redshift degeneracy~\cite{Ezquiaga:2022zkx,Mastrogiovanni:2021wsd}.
Both sources of redshift information are combined with the luminosity-distance posterior from GW data within a hierarchical Bayesian framework, in which the likelihood of multiple GW events can be written as \begin{equation}
\mathcal{L}_{\rm GW}(\Theta)
\propto
e^{-N_{\rm exp}(\Theta)}
\prod_{i=1}^{N_{\rm obs}}
\left[
\frac{T_{\rm obs}}{N_{s,i}}
\sum_{j=1}^{N_{s,i}}
w_{ij}(\Theta)
\right],
\end{equation}
where $\Theta$ denotes the set of cosmological parameters and compact-binary population hyperparameters, $N_{\rm exp}(\Theta)$ is the expected number of detected events for a given model and accounts for selection effects, $N_{\rm obs}$ is the number of observed GW events, $T_{\rm obs}$ is the observing time, $N_{s,i}$ is the number of posterior samples used for the $i$-th event, and $w_{ij}(\Theta)$ is the importance-sampling reweighting factor for the $j$-th posterior sample of the $i$-th event, defined as the ratio of the astrophysical merger rate $dN/(dt\,d\theta)$ evaluated at the sample parameters to the prior $\pi_{\rm PE}$ used in the original parameter estimation~\cite{Mastrogiovanni:2023zbw}.

In practice, we compute the GW likelihood using the public package \texttt{icarogw}\footnote{\url{https://github.com/simone-mastrogiovanni/icarogw}}~\cite{Mastrogiovanni:2023zbw}. The \texttt{icarogw} package implements a hierarchical Bayesian inference framework for joint analyses of GW standard sirens, population models, and cosmological parameters. We use \texttt{icarogw} version 2.0.3, consistent with the version adopted in the official GWTC-4 dark siren analysis~\cite{LIGOScientific:2025jau}.

For the GW population model, we adopt both the FullPop-4.0 and MLTP models. The FullPop-4.0 model includes a power-law component at low masses, a power-law structure in the black hole mass regime, two Gaussian peaks, and a dip feature around the gap region in the single-object mass distribution. This flexible parameterization allows BNS, NSBH, and BBH events to be described within a unified framework. The primary-mass distribution takes the form
\begin{equation}
\begin{aligned}
p_{\rm S}(m_1 | \Lambda) \propto{}&
p(m_1 | \Lambda)\,
S_{\rm h}(m_1 | m_{\min},\delta_m^{\min}) \\
&\times S_{\rm l}(m_1 | m_{\max},\delta_m^{\max}) \\
&\times S_{\rm n}\!\left(
m_1 | m_d^{\rm low},\delta_d^{\min},
m_d^{\rm high},\delta_d^{\max}
\right) ,
\end{aligned}
\end{equation}
where $p(m_1|\Lambda)$ denotes the underlying primary-mass distribution before applying boundary smoothing and mass-gap corrections. The function $S_{\rm h}(m_1 | m_{\min},\delta_m^{\min})$ is the high-pass smoothing function at the low-mass end, $S_{\rm l}(m_1 | m_{\max},\delta_m^{\max})$ is the low-pass smoothing function at the high-mass end, and $S_{\rm n}(m_1 | m_d^{\rm low},\delta_d^{\min},m_d^{\rm high},\delta_d^{\max})$ is the mass-gap correction function. The secondary-mass distribution adopts a similar single-object mass distribution form; the full parameterization can be found in Ref.~\cite{LIGOScientific:2025jau}.

The MLTP model is applied only to BBH events. Its primary-mass distribution consists of a power-law component and two Gaussian peaks, while the secondary-mass distribution is described through a power-law distribution in the mass ratio~\cite{LIGOScientific:2025jau}.

When combining GW standard sirens with strong gravitational lensing systems, the GW luminosity distance needs to be converted into the dimensionless distance required by the strong-lensing framework:
\begin{equation}
d(z)=\frac{H_0 D_L(z)}{c(1+z)}.
\end{equation}
Following previous studies~\cite{Collett:2019hrr,Qi:2020rmm,Wei:2020suh}, we reconstruct a continuous distance-redshift relation by parameterizing $d(z)$ at low redshift with an $n$-th order polynomial:
\begin{equation}
d(z)=z+\sum_{k=2}^{n} a_{k-1} z^k,
\end{equation}
which satisfies the physically motivated boundary conditions $d(0)=0$ and $d'(0)=1$ by construction. The optimal polynomial order is determined by model selection criteria in Section~\ref{sec:results}.

In the joint analysis, we combine the GW and strong-lensing likelihoods as
\begin{equation}
\ln \mathcal{L} = \ln \mathcal{L}_{\rm SGL} + \ln \mathcal{L}_{\rm GW},
\end{equation}
where $\mathcal{L}_{\rm GW}$ is the likelihood for the GWTC-4 standard siren events and $\mathcal{L}_{\rm SGL}$ is the likelihood for the SGLTD systems.
We employ the public Bayesian inference framework \texttt{bilby}~\cite{Ashton:2018jfp} together with the \texttt{nessai} sampler~\cite{michael_j_williams_2024_10965503,Williams:2021qyt} to jointly sample cosmological parameters, GW population parameters, and strong-lensing-related parameters.

\subsection{Data}\label{sec:data}

\subsubsection{GW data}

Following the official GWTC-4 analysis~\cite{LIGOScientific:2025jau}, we select GW events detected during the O1--O4a observing runs with a false-alarm rate (FAR) below $0.25\,{\rm yr^{-1}}$, taking the lowest FAR among all search pipelines, yielding a total of 142 events at redshifts $z\lesssim 1$: 137 binary black hole (BBH) mergers, 3 neutron star-black hole (NSBH) mergers, and 2 binary neutron star (BNS) mergers.
For the FullPop-4.0 analysis, we use all 142 compact binary coalescence (CBC) events; for the MLTP analysis, which describes only the BBH population, we use 137 BBH events together with the bright siren GW170817.
We use $K$-band galaxies from the GLADE+ catalog~\cite{Dalya:2018cnd,Dalya:2021ewn} and assume that their absolute magnitudes follow a Schechter function~\cite{Kochanek:2000im}.

We adopt a uniform prior $H_0 \in \mathcal{U}(10,200)\ {\rm km\,s^{-1}\,Mpc^{-1}}$, and uniform priors $\mathcal{U}(-1,1)$ for all polynomial coefficients $a_{k-1}$ ($k=2,\dots,n$).
The prior ranges of the FullPop-4.0 population hyperparameters are summarized in Table~\ref{tab:gw_priors}, while the priors for the MLTP population model follow the official GWTC-4 analysis~\cite{LIGOScientific:2025jau}.
Song et al.~\cite{Song:2025ddm} found a strong positive degeneracy between $H_0$ and $\Omega_K$ in the distance-sum-rule framework using GWTC-3 data, indicating that the current generation of standard siren data cannot yet place meaningful constraints on $\Omega_K$. For simplicity, we fix the spatial curvature parameter to $\Omega_K=0$ in this work.

\begin{table*}[t]
\centering
\caption{Summary of the priors adopted for the hyperparameters of the FullPop-4.0 mass population model. U (LU) denotes uniform (log-uniform) prior.}
\label{tab:gw_priors}
\begin{tabular}{ccc}
\hline\hline
Parameter & Prior & Description \\
\hline
$\alpha_1$ & $\mathcal{U}(-4,12)$ & Power-law slope describing the low-mass component of the primary-mass distribution \\
$\alpha_2$ & $\mathcal{U}(-4,12)$ & Power-law slope governing the high-mass portion of the primary-mass distribution \\
$\beta_1$ & $\mathcal{U}(-4,12)$ & Power-law index for the lower-mass regime of the secondary component \\
$\beta_2$ & $\mathcal{U}(-4,12)$ & Power-law index for the higher-mass regime of the secondary component \\
$m_{\min}$ & $\mathcal{U}(0.4,1.4)$ & Lower bound of the compact-object mass spectrum \\
$m_{\max}$ & $\mathcal{U}(50,200)$ & Upper bound of the compact-object mass spectrum \\
$\delta_m^{\min}$ & $\mathcal{LU}(10^{-2},1)$ & Width parameter controlling the smoothing of the low-mass cutoff \\
$\delta_m^{\max}$ & $\mathcal{LU}(10^{-3},1)$ & Smoothing scale applied near the upper-mass boundary \\
$\mu_g^{\rm low}$ & $\mathcal{U}(5,150)$ & Mean value of the first Gaussian component in the mass spectrum \\
$\sigma_g^{\rm low}$ & $\mathcal{U}(0.4,5)$ & Standard deviation of the first Gaussian component \\
$\mu_g^{\rm high}$ & $\mathcal{U}(5,150)$ & Mean value of the second Gaussian component \\
$\sigma_g^{\rm high}$ & $\mathcal{U}(0.4,10)$ & Standard deviation of the second Gaussian component \\
$\lambda_g$ & $\mathcal{U}(0,1)$ & Fraction of the population described by Gaussian features \\
$\lambda_g^{\rm low}$ & $\mathcal{U}(0,1)$ & Relative contribution of the first Gaussian component \\
$m_d^{\rm low}$ & $\mathcal{U}(1.5,3)$ & Lower boundary of the dip feature in the mass distribution \\
$m_d^{\rm high}$ & $\mathcal{U}(5,9)$ & Upper boundary of the dip feature in the mass distribution \\
$\delta_d^{\min}$ & $\mathcal{LU}(0.01,2)$ & Smoothing scale on the low-mass side of the dip \\
$\delta_d^{\max}$ & $\mathcal{LU}(0.01,2)$ & Smoothing scale on the high-mass side of the dip \\
$A$ & $\mathcal{U}(0,1)$ & Depth (amplitude) of the dip feature \\
\hline\hline
\end{tabular}
\end{table*}

\subsubsection{Strong lensing data}

\begin{figure}[tbp]
\centering
\includegraphics[width=0.45\textwidth]{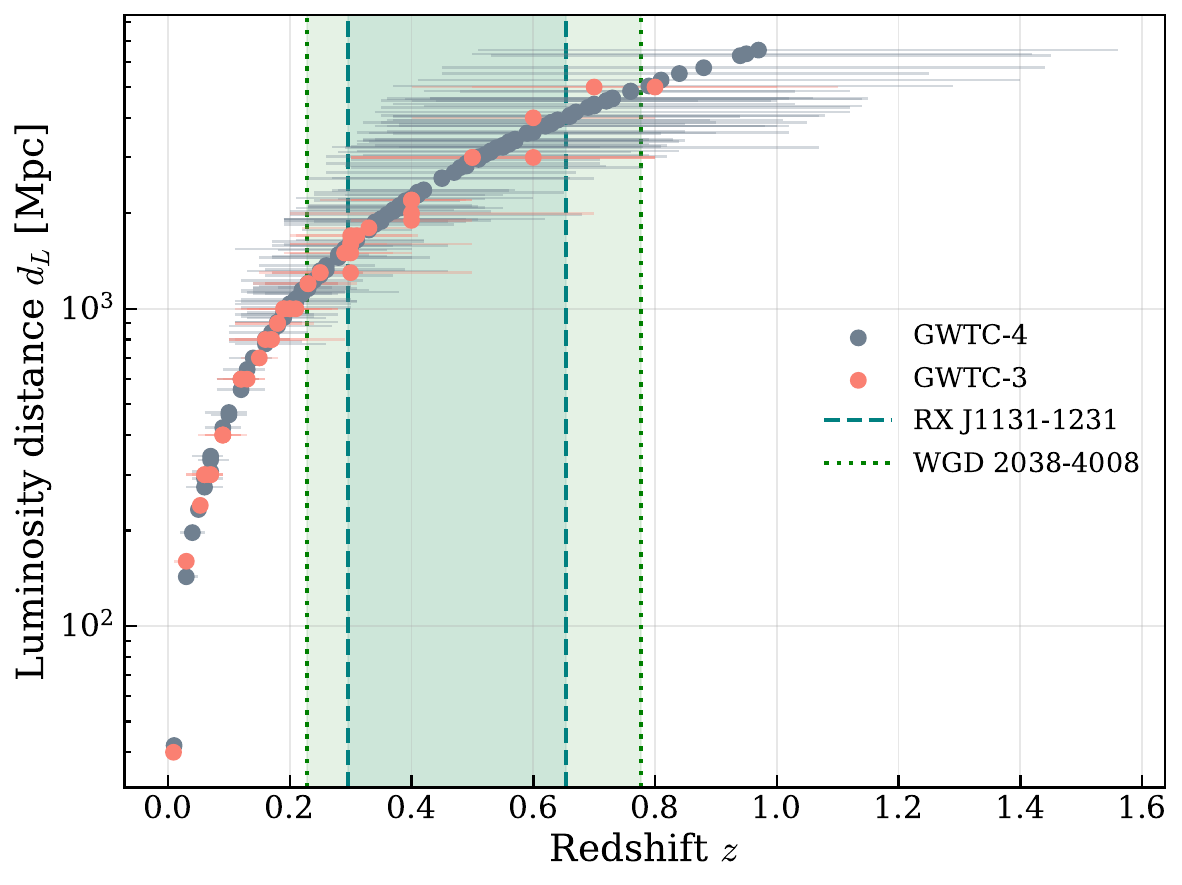}
\caption{Distribution of the GW standard siren samples and the selected SGLTD systems in the redshift-distance plane. The gray points denote the GWTC-4 events, while the orange-red points denote the GWTC-3 events. The redshift ranges of the GW events are obtained by converting the luminosity-distance posteriors from the GW data assuming the $\Lambda$CDM model with Planck 2018 cosmological parameters~\cite{Planck:2018vyg}. The two different green vertical lines and shaded regions mark the lens and source redshift ranges of the two SGLTD systems, RX J1131-1231 and WGD 2038-4008.}
\label{fig:gw_lens_data}
\end{figure}

We use two SGLTD systems from the latest TDCOSMO2025 lens sample: RX J1131-1231~\cite{Suyu:2013kha,H0LiCOW:2019xdh} and WGD 2038-4008~\cite{DES:2019fny}.
Figure~\ref{fig:gw_lens_data} shows the distribution of the GW standard siren samples and the selected strong lensing systems in the redshift-luminosity-distance plane.
The spectroscopic redshifts of both the lens galaxies and the background sources of the two selected lens systems lie within the redshift coverage of the GWTC-4 data.
This ensures that the distance quantities required by the strong lensing systems can be reconstructed directly within the redshift range constrained by the standard sirens.
For RX J1131-1231, the lens and source redshifts are $z_l=0.295$~\cite{Sluse:2003iy} and $z_s=0.654$~\cite{Sluse:2007cn}, respectively.
For WGD 2038-4008, the corresponding redshifts are $z_l=0.228$~\cite{DES:2017bks} and $z_s=0.777$~\cite{DES:2020ohe}, respectively.

In this work, we consider a configuration that uses only the two time-delay lens systems RX J1131-1231 and WGD 2038-4008, without including external lens samples.
We refer to this configuration as the TDCOSMO2025-only analysis.
We also consider a configuration that includes the external non-time-delay lens samples SLACS and SL2S, which is referred to as the TDCOSMO2025-SLACS-SL2S analysis.
The parameters and prior ranges used in the strong lensing analysis are summarized in Table~\ref{tab:td_priors}.

In addition, to compare the impact of different MST treatments on the results, we also perform an H0LiCOW-based analysis using only RX J1131-1231, following the setup adopted by Song et al.~\cite{Song:2025ddm}.

\begin{table*}[t]
\centering
\caption{Prior distributions adopted for the parameters associated with the time-delay lensing model.
$\mathcal{U}$ ($\mathcal{LU}$) denotes a uniform (log-uniform) prior distribution.}
\label{tab:td_priors}
\begin{tabular}{ccc}
\hline\hline
Parameter & Prior & Description \\
\hline

$\lambda_{\rm int,0}$ & $\mathcal{U}(0.5,1.5)$
& Central value of the internal mass-sheet transformation (MST) parameter for the lens population \\

$\alpha_{\lambda}$ & $\mathcal{U}(-1,1)$
& Coefficient describing the dependence of $\lambda_{\rm int}$ on the ratio $R_{\rm eff}/\theta_{\rm E}$ \\

$\sigma(\lambda_{\rm int})$ & $\mathcal{LU}(10^{-3},0.5)$
& Intrinsic dispersion of the internal MST parameter across the lens sample \\

$\gamma_{\rm pl,i}$ & $\mathcal{U}(1.5,2.5)$
& Power-law slope characterizing the mass density profile of each individual lens \\

$\langle \sigma_t / \sigma_r \rangle$ & $\mathcal{U}(0.87,1.12)$
& Population-averaged stellar velocity anisotropy assuming a spatially constant anisotropy model \\

$\sigma(\sigma_t/\sigma_r)$ & $\mathcal{LU}(0.01,1)$
& Scatter in the anisotropy ratio $\sigma_t/\sigma_r$ among different lens systems \\

\hline\hline
\end{tabular}
\end{table*}

\section{Results and discussions}\label{sec:results}

To determine the optimal polynomial order for distance reconstruction, we compare second-, third-, and fourth-order polynomials under both the FullPop-4.0 and MLTP GW population models, with the population hyperparameters fixed during the test. The results are summarized in Table~\ref{tab:aic_bic}.

For each candidate, we compute the maximum likelihood $\mathcal{L}_{\rm max}$ from the GW data and evaluate the Akaike Information Criterion (AIC)~\cite{Akaike:1974vps} and the Bayesian Information Criterion (BIC)~\cite{Schwarz:1978tpv}:
\begin{equation}
{\rm AIC} = -2 \ln \mathcal{L}_{\rm max} + 2 k,
\end{equation}
\begin{equation}
{\rm BIC} = -2 \ln \mathcal{L}_{\rm max} + k \ln N,
\end{equation}
where $k$ is the number of free parameters and $N$ is the number of GW events. The relative performance of each model is quantified by
\begin{equation}
\Delta {\rm AIC}_i = {\rm AIC}_i - {\rm AIC}_{\rm min},
\end{equation}
\begin{equation}
\Delta {\rm BIC}_i = {\rm BIC}_i - {\rm BIC}_{\rm min}.
\end{equation}

As shown in Table~\ref{tab:aic_bic}, the AIC favors the third-order polynomial under both population models. The BIC minimum occurs at second order owing to its stronger complexity penalty; however, the $\Delta$BIC between the second- and third-order polynomials is small ($\approx 0.95$ for FullPop-4.0 and $\approx 0.14$ for MLTP), providing no strong evidence against the third-order model. The fourth-order polynomial does not improve the fit sufficiently to offset the additional complexity penalty and is disfavored by both criteria. Based on these results and previous studies, we adopt the third-order polynomial $d(z)=z+a_1 z^2+a_2 z^3$ as our baseline distance reconstruction model.

\begin{table*}[t]
\centering
\caption{
AIC and BIC comparison for different polynomial orders under the FullPop-4.0 and MLTP population models.
The definitions of $\Delta{\rm AIC}$ and $\Delta{\rm BIC}$ are given in the text.
}
\label{tab:aic_bic}
\renewcommand{\arraystretch}{1.35}
\setlength{\tabcolsep}{7pt}
\begin{tabular}{ccccccccc}
\hline\hline
Population model & Order & $a_1$ & $a_2$ & $a_3$ & BIC & AIC & $\Delta{\rm BIC}$ & $\Delta{\rm AIC}$ \\
\hline
\multirow{3}{*}{FullPop-4.0}
& Second
& $-0.11^{+0.03}_{-0.01}$
& --
& --
& 4387.239
& 4381.328
& --
& 2.002 \\
\cline{2-9}

& Third
& $-0.36^{+0.11}_{-0.13}$
& $0.10^{+0.09}_{-0.06}$
& --
& 4388.194
& 4379.326
& 0.954
& -- \\
\cline{2-9}

& Fourth
& $0.10^{+0.27}_{-0.31}$
& $-0.54^{+0.41}_{-0.31}$
& $0.22^{+0.14}_{-0.14}$
& 4392.951
& 4381.128
& 5.712
& 1.802 \\
\hline

\multirow{3}{*}{MLTP}
& Second
& $-0.11^{+0.03}_{-0.01}$
& --
& --
& 4217.980
& 4212.140
& --
& 2.781 \\
\cline{2-9}

& Third
& $-0.39^{+0.11}_{-0.15}$
& $0.11^{+0.10}_{-0.06}$
& --
& 4218.119
& 4209.359
& 0.139
& -- \\
\cline{2-9}

& Fourth
& $0.09^{+0.27}_{-0.31}$
& $-0.56^{+0.40}_{-0.31}$
& $0.24^{+0.14}_{-0.15}$
& 4222.910
& 4211.230
& 4.930
& 1.871 \\
\hline\hline
\end{tabular}
\end{table*}

We perform a joint analysis of the GWTC-4 standard siren data and the SGLTD data.
Within the third-order polynomial distance reconstruction framework, we jointly constrain $H_0$, the distance reconstruction parameters, the GW source population hyperparameters, and the strong-lensing-related parameters.
The constraints on $H_0$ and their relative precisions are summarized in Table~\ref{tab:h0_results}.
Under the FullPop-4.0 population model with the TDCOSMO2025-only lensing configuration, we obtain $H_0 = 83.78^{+12.53}_{-10.23}\ {\rm km\,s^{-1}\,Mpc^{-1}}$, corresponding to a relative precision of $13.58\%$.

To examine the impact of the GW source population model, we also perform the analysis under the MLTP model using 137 BBH events together with the bright siren GW170817, obtaining $H_0 = 94.21^{+11.41}_{-13.14}\ {\rm km\,s^{-1}\,Mpc^{-1}}$ with a relative precision of $13.03\%$.
Compared with the FullPop-4.0 result, the MLTP central value shifts notably higher, while the two models yield comparable relative precisions ($13.03\%$ versus $13.58\%$).
The difference in the central value reflects the mass-redshift degeneracy inherent in dark siren analyses: since the source-frame masses are related to the detector-frame masses by $m_{\rm source} = m_{\rm det}/(1+z)$, different population models with different mass-spectrum parameterizations and event samples anchor the redshift inference differently, leading to shifts in the inferred $H_0$~\cite{LIGOScientific:2025jau}.This interpretation is supported by recent spectral-siren studies of GWTC-4 BBH events, which show that at current sensitivity, redshift evolution of the BBH mass spectrum has a negligible impact on \(H_0\), while the dominant systematic is the adopted redshift-independent mass spectrum parameterization~\cite{Agapito:2026kak}.
Notably, the official GWTC-4 dark siren analysis finds that FullPop-4.0, which extends the mass spectrum to NS-containing events and introduces additional characteristic mass scales such as the NS mass peaks and the BH-NS mass gap, provides significantly tighter $H_0$ constraints than MLTP~\cite{LIGOScientific:2025jau}.
However, this advantage is not reflected in our joint analysis, where the two models give nearly identical precisions, indicating that the final $H_0$ precision in our framework is dominated by MST-related uncertainties on the strong-lensing side rather than by the GW-side redshift inference capability.

Next, under the GWTC-4 + FullPop-4.0 setting, we include the SLACS and SL2S external non-time-delay lens samples, extending the lensing configuration to TDCOSMO2025+SLACS+SL2S.
The corresponding result is $H_0 = 88.13^{+12.80}_{-10.63}\ {\rm km\,s^{-1}\,Mpc^{-1}}$, with a relative precision of $13.3\%$.
Compared with the TDCOSMO2025-only result ($13.58\%$), the precision is essentially unchanged, while the central value shifts higher.
This is consistent with the TDCOSMO2025 analysis~\cite{TDCOSMO:2025dmr}, where the addition of external lens samples also yields only a modest improvement in $H_0$ precision.
As discussed in Ref.~\cite{TDCOSMO:2025dmr}, the limited improvement is driven not by data quality but by the intrinsic scatter in $\lambda_{\rm int}$ among the SLACS lenses, which places a floor on the achievable precision. Furthermore, the SLACS sample prefers a slightly higher $\lambda_{\rm int}$ than the TDCOSMO time-delay lenses alone, which drives the central value of $H_0$ upward when the samples are combined.

We note that a direct comparison with the GWTC-3 result of Song et al.~\cite{Song:2025ddm} ($H_0 = 70.40^{+8.03}_{-5.60}\ {\rm km\,s^{-1}\,Mpc^{-1}}$, precision $9.68\%$, using the TDCOSMO-4 framework~\cite{Birrer:2020tax} with a single lens RX~J1131-1231) is not straightforward, as the two analyses differ in the GW data, lens samples, population models, and the strong-lensing framework.
In particular, the TDCOSMO2025 framework~\cite{TDCOSMO:2025dmr} supersedes TDCOSMO-4 by incorporating substantially improved stellar kinematic data from JWST, Keck, and VLT, and replacing the SDSS-based SLACS kinematics, which were shown to suffer from systematic biases, with high-quality IFU measurements.
While these improvements yield more reliable constraints, the higher-quality kinematic data also reveal significant intrinsic scatter in $\lambda_{\rm int}$ among the SLACS lenses that was undetectable with the lower signal-to-noise SDSS spectra used in TDCOSMO-4, inherently limiting the achievable $H_0$ precision~\cite{TDCOSMO:2025dmr}.
On the GW side, the more flexible FullPop-4.0 population model used in GWTC-4 introduces additional parameter degeneracies compared with the simpler Power Law + Peak model adopted in the GWTC-3 analysis, which offsets part of the statistical gain from the larger event sample~\cite{LIGOScientific:2025jau}.
The broader $H_0$ constraint in our analysis therefore results from methodological updates on both sides that improve robustness at the cost of nominal precision.

\begin{table*}[t]
\centering
\caption{
Summary of the $H_0$ constraints obtained from different combinations of GW standard siren data, SGLTD data, population models, and MST treatment methods.
$H_0$ is given in units of $\mathrm{km\,s^{-1}\,Mpc^{-1}}$.
The precision is defined as $\Delta H_0/H_0$, where $\Delta H_0$ is taken as the average of the upper and lower $68.3\%$ credible intervals. The two GWTC-3 rows in the table are taken from Song et al.~\cite{Song:2025ddm} and are used in this work as a comparison with the GWTC-4 results.}

\label{tab:h0_results}

\renewcommand{\arraystretch}{1.65}
\setlength{\tabcolsep}{6pt}
\setcellgapes{4pt}
\makegapedcells

\begin{tabular*}{\textwidth}{@{\extracolsep{\fill}}ccccc@{}}
\hline\hline
\makecell[c]{GW data}
& \makecell[c]{SGLTD data}
& \makecell[c]{Population model}
& \makecell[c]{MST treatment method}
& \makecell[c]{$H_0$ constraint and precision} \\
\hline

\makecell[c]{GWTC-4 \\ [2pt]142 CBCs}
& \makecell[c]{RX J1131-1231 \\ WGD 2038-4008}
& \makecell[c]{FullPop-4.0}
& \makecell[c]{TDCOSMO2025-only}
& \makecell[c]{$83.78^{+12.53}_{-10.23}$ $(13.58\%)$} \\
\hline

\makecell[c]{GWTC-4 \\ [2pt]137 BBHs \\ [2pt]GW170817}
& \makecell[c]{RX J1131-1231 \\ [2pt]WGD 2038-4008}
& \makecell[c]{MLTP}
& \makecell[c]{TDCOSMO2025-only}
& \makecell[c]{$94.21^{+11.41}_{-13.14}$ $(13.03\%)$} \\
\hline

\makecell[c]{GWTC-4 \\ [2pt]142 CBCs}
& \makecell[c]{RX J1131-1231 \\ [2pt]WGD 2038-4008}
& \makecell[c]{FullPop-4.0}
& \makecell[c]{TDCOSMO2025-SLACS-SL2S}
& \makecell[c]{$88.13^{+12.80}_{-10.63}$ $(13.3\%)$} \\
\hline

\makecell[c]{GWTC-3 \\ [2pt]47 CBCs}
& \makecell[c]{RX J1131-1231}
& \makecell[c]{Power Law + Peak}
& \makecell[c]{TDCOSMO}
& \makecell[c]{$70.40^{+8.03}_{-5.60}$ $(9.68\%)$} \\
\hline

\makecell[c]{GWTC-4 \\ [2pt]142 CBCs}
& \makecell[c]{RX J1131-1231}
& \makecell[c]{FullPop-4.0}
& \makecell[c]{H0LiCOW}
& \makecell[c]{$75.42^{+3.74}_{-4.66}$ $(5.57\%)$} \\
\hline

\makecell[c]{GWTC-3 \\ [2pt]47 CBCs}
& \makecell[c]{RX J1131-1231}
& \makecell[c]{Power Law + Peak}
& \makecell[c]{H0LiCOW}
& \makecell[c]{$73.22^{+5.95}_{-5.43}$ $(7.77\%)$} \\

\hline\hline
\end{tabular*}
\end{table*}

\begin{figure}[!t]
\centering
\includegraphics[width=0.48\textwidth]{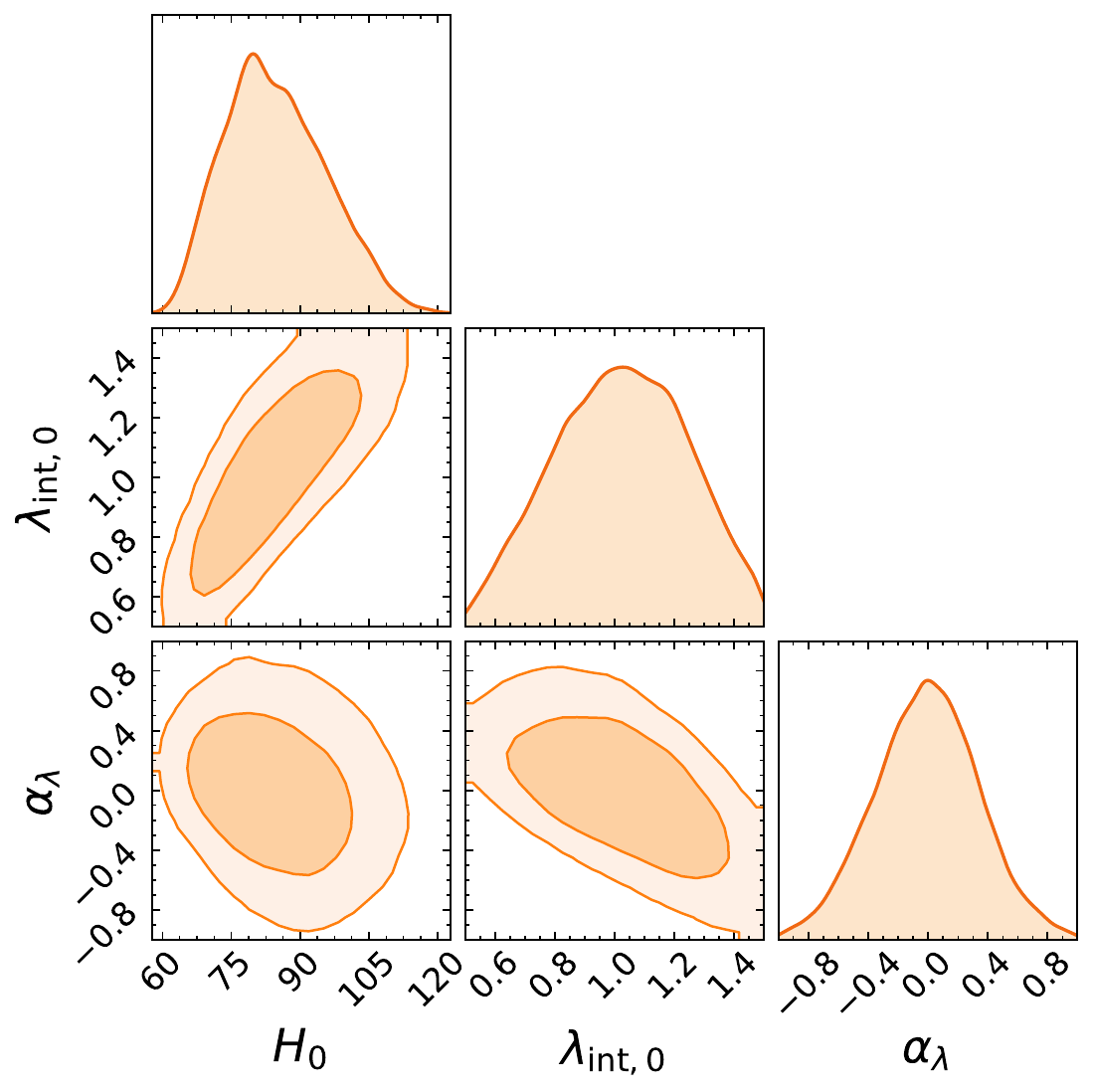}
\caption{Joint posterior distributions of $H_0$, the population mean of the internal MST parameter $\lambda_{\rm int,0}$, and the slope $\alpha_{\lambda}$ describing the dependence of $\lambda_{\rm int}$ on $R_{\rm eff}/\theta_{\rm E}$, obtained from the GWTC-4 + TDCOSMO2025-only analysis with third-order polynomial distance reconstruction. The inner and outer contours enclose the $68.3\%$ and $90\%$ credible regions, respectively.}
\label{fig:corner_H0}
\end{figure}

In addition, we perform a comparison by replacing the TDCOSMO2025 treatment with the H0LiCOW method, using only RX~J1131-1231 while keeping the GWTC-4 data and distance reconstruction unchanged.
This yields $H_0 = 75.42^{+3.74}_{-4.66}\ {\rm km\,s^{-1}\,Mpc^{-1}}$ with a relative precision of $5.57\%$, significantly tighter than the TDCOSMO2025-only result ($13.58\%$).
This contrast is expected: the $H_0$ constraint in this framework is primarily driven by the SGLTD data~\cite{Song:2025ddm}, making the choice of MST treatment the dominant factor in determining the final precision.
The H0LiCOW method breaks the MST by assuming specific parametric mass density profiles for the lens galaxy~\cite{H0LiCOW:2019xdh}, whereas the TDCOSMO2025 hierarchical approach parametrizes $\lambda_{\rm int}$ at the population level and relies solely on stellar kinematics to constrain the degeneracy~\cite{TDCOSMO:2025dmr}, propagating correspondingly larger uncertainties into the $H_0$ posterior.
This propagation is directly visible in Figure~\ref{fig:corner_H0}: $H_0$ exhibits a clear positive degeneracy with $\lambda_{\rm int,0}$, the population mean of the internal MST parameter, and also correlates with the slope $\alpha_{\lambda}$, illustrating how the MST-related parameter uncertainties broaden the $H_0$ constraint under the TDCOSMO2025 framework.
Conversely, when the lensing-side treatment is held fixed at H0LiCOW, the GW-side improvement becomes apparent: our GWTC-4 + H0LiCOW precision ($5.57\%$) improves upon the GWTC-3 + H0LiCOW result of Song et al.~\cite{Song:2025ddm} ($7.77\%$) by approximately 2 percentage points, reflecting both the larger event sample and the more comprehensive FullPop-4.0 population model.

\begin{figure*}[t]
\centering
\includegraphics[width=0.9\textwidth]{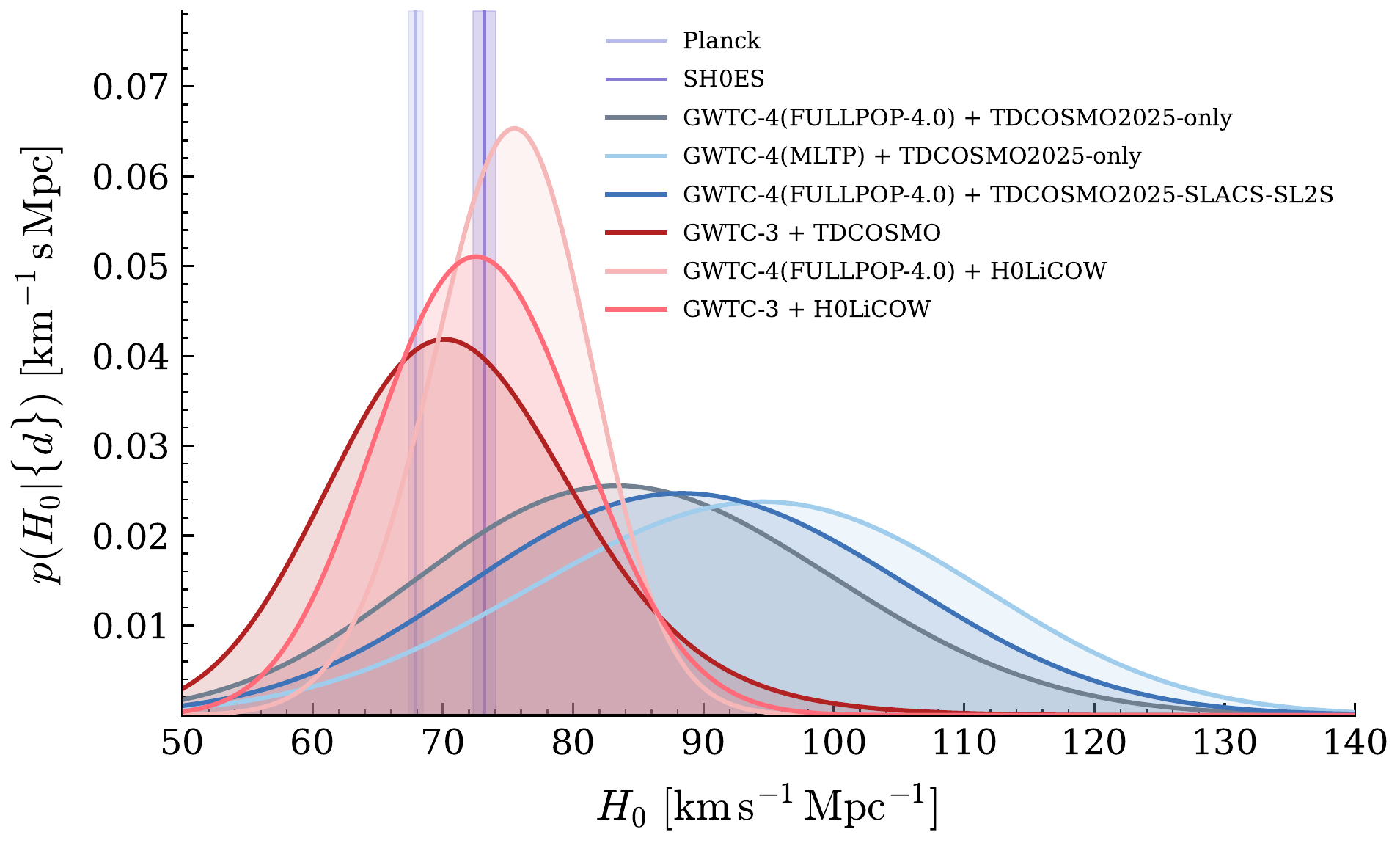}
\caption{One-dimensional posterior distributions of $H_0$ obtained from different data combinations and strong-lensing treatments.
The blue-gray curve represents the result from GWTC-4(FullPop-4.0) + TDCOSMO2025-only, the light-blue curve represents GWTC-4(MLTP) + TDCOSMO2025-only, and the dark-blue curve represents GWTC-4(FullPop-4.0) + TDCOSMO2025-SLACS-SL2S.
The dark-red curve represents GWTC-3 + TDCOSMO, the light-pink curve represents GWTC-4(FullPop-4.0) + H0LiCOW, and the pink curve represents GWTC-3 + H0LiCOW.
The vertical shaded regions indicate the $H_0$ reference intervals from Planck and SH0ES, respectively.}
\label{fig:H0_posteriors}
\end{figure*}

Finally, Figure~\ref{fig:H0_posteriors} summarizes the one-dimensional $H_0$ posteriors from all analyses in Table~\ref{tab:h0_results}.
The three TDCOSMO2025-based results (FullPop-4.0, MLTP, and FullPop-4.0 + SLACS + SL2S) all exhibit broad and comparable posteriors, confirming that the addition of external lens samples does not significantly tighten the constraint under the current hierarchical MST framework.
In contrast, the H0LiCOW treatment yields substantially more concentrated posteriors, with the GWTC-4 + H0LiCOW result being the tightest overall.
Given the current precision, all results remain consistent with both the Planck and SH0ES reference intervals and cannot yet distinguish between the two.

\section{Conclusions}\label{sec:conclusions}

In this work, we present a cosmological-model-independent constraint on $H_0$ by combining GWTC-4 GW standard siren data with TDCOSMO2025 SGLTD systems under the FLRW distance-sum-rule framework. Our main findings are as follows.

Under the FullPop-4.0 population model with the TDCOSMO2025-only lensing configuration, we obtain $H_0 = 83.78^{+12.53}_{-10.23}\ {\rm km\,s^{-1}\,Mpc^{-1}}$ ($13.58\%$). The MLTP model yields $H_0 = 94.21^{+11.41}_{-13.14}\ {\rm km\,s^{-1}\,Mpc^{-1}}$ ($13.03\%$), with a higher central value but comparable precision. The shift in central value reflects the mass-redshift degeneracy: different population models anchor the redshift inference differently. The fact that both models give nearly identical precisions, despite the official GWTC-4 analysis showing a clear advantage for FullPop-4.0 in pure dark siren constraints, indicates that the $H_0$ precision in our framework is dominated by the MST-related uncertainties on the strong-lensing side.

Including the SLACS and SL2S external non-time-delay lens samples yields $H_0 = 88.13^{+12.80}_{-10.63}\ {\rm km\,s^{-1}\,Mpc^{-1}}$ ($13.3\%$), with the precision essentially unchanged and the central value shifted higher. This is consistent with the TDCOSMO2025 analysis, which finds that the improvement from external samples is limited by the intrinsic scatter in $\lambda_{\rm int}$ among the SLACS lenses.

Replacing the TDCOSMO2025 treatment with the H0LiCOW method gives $H_0 = 75.42^{+3.74}_{-4.66}\ {\rm km\,s^{-1}\,Mpc^{-1}}$ ($5.57\%$), confirming that the choice of MST treatment is the dominant factor governing the $H_0$ precision, as the constraint is primarily driven by the SGLTD data. When the lensing-side treatment is held fixed at H0LiCOW, the GW-side improvement becomes apparent: the GWTC-4 precision ($5.57\%$) improves upon the GWTC-3 result ($7.77\%$) by approximately 2 percentage points, reflecting the larger event sample and the more comprehensive FullPop-4.0 population model. At the current level of precision, all results are consistent with both the Planck and SH0ES values and cannot yet distinguish between the two.

Looking ahead, the precision of this model-independent approach can be improved from both sides. On the GW side, more high-redshift dark siren events from future observing runs will strengthen the distance-redshift calibration and, crucially, extend the redshift coverage to accommodate additional SGLTD systems whose lens or source redshifts lie beyond the reach of the current sample. Recent GWTC-4 dark-siren analyses with deep photometric galaxy catalogs further show that improved catalog depth can enhance the constraining power, although redshift-dependent catalog features and selection effects must be carefully controlled~\cite{McMahon:2026nhi}. On the lensing side, a larger number of SGLTD systems, higher-quality spatially resolved stellar kinematics, and better-understood external lens samples will help reduce the MST-related uncertainties that currently dominate the error budget. The synergy between these two improvements is essential: high-redshift GW events enable the inclusion of more SGLTD systems, which in turn provide tighter constraints on the MST parameters and $H_0$.

\begin{acknowledgments}

This work was supported by the National Natural Science Foundation of China
(Grants Nos. 12473001, 12575049, and 12533001), the National SKA Program of China
(Grants Nos. 2022SKA0110200 and 2022SKA0110203), the China Manned Space Program
(Grant No. CMS-CSST-2025-A02), and the 111 Project (Grant No. B16009).

\end{acknowledgments}

\bibliography{ref}

\end{document}